\documentclass[11pt]{article}
\usepackage[utf8]{inputenc}
\usepackage{amsfonts}
\usepackage{amsmath}
\usepackage{amssymb}
\usepackage{indentfirst}
\usepackage{graphicx}
\usepackage[colorlinks]{hyperref}
\usepackage{cite}
\usepackage{array}
\usepackage{microtype}
\usepackage{soul}
\usepackage[normalem]{ulem}
\usepackage{cancel}
\usepackage{soul}
\usepackage{xspace}
\usepackage{colortbl}
\usepackage{float}
\usepackage{caption}
\numberwithin{equation}{section}
\allowdisplaybreaks

\makeatletter

\begin{document}

\begin{titlepage}
\vspace{3cm}

\baselineskip=24pt

\begin{center}
\textbf{\LARGE Non-Relativistic Chern-Simons Supergravity with Torsion}
\par\end{center}{\LARGE \par}

\begin{center}
	\vspace{1cm}
    \textbf{Francisco Barriga}$^{1}$,
	\textbf{Patrick Concha}$^{2,3}$,
	\textbf{Nelson Merino}$^{4,5}$,
	\textbf{Evelyn Rodríguez}$^{2,3}$,
	\small
	\\[5mm]
   $^{1}$ \textit{Departamento de Física,}\\ \textit{Universidad del Bío-Bío, Casilla 5-C, Concepción, Chile. } \\[2mm]
    $^{2}$\textit{Departamento de Matemática y Física Aplicadas, }\\
	\textit{ Universidad Católica de la Santísima Concepción, }\\
\textit{ Alonso de Ribera 2850, Concepción, Chile.}
    \\[2mm]
    $^{3}$\textit{Grupo de Investigación en Física Teórica, GIFT, }\\
	\textit{ Universidad Católica de la Santísima Concepción, }\\
\textit{ Alonso de Ribera 2850, Concepción, Chile.}
\\[2mm]
    $^{4}$\textit{Instituto de Ciencias Exactas y Naturales,}\\\textit{Universidad Arturo Prat, Playa Brava 3265, 1111346, Iquique, Chile.}
    \\[2mm]
    $^{5}$\textit{Facultad de Ciencias, Universidad Arturo Prat,}\\\textit{Avenida Arturo Prat Chacón 2120, 1110939, Iquique, Chile.}
	 \\[5mm]
	\footnotesize
    \texttt{francisco.barriga2401@alumnos.ubiobio.cl}
	\texttt{patrick.concha@ucsc.cl},
	\texttt{nemerino@unap.cl},
	\texttt{erodriguez@ucsc.cl},
	\par\end{center}
\vskip 26pt
\begin{abstract}

In this work, we construct a three-dimensional non-relativistic Chern--Simons supergravity theory with both curvature and torsion within the Mielke--Baekler framework. We show that a consistent non-relativistic supergravity formulation requires starting from a $\mathcal{N}=2$ supersymmetric extension of the Mielke--Baekler algebra and implementing a non-relativistic expansion via the semigroup expansion method, rather than a naive contraction. This procedure allows one to overcome the usual difficulties of non-relativistic supergravity constructions, ensuring closure of the superalgebra and the existence of a non-degenerate invariant bilinear form. The resulting model is characterized by two parameters $(p,q)$, which interpolate between different non-relativistic supergravity theories, including the extended Bargmann, Newton--Hooke, and torsional models. Our results provide a unified framework for non-relativistic supergravity with torsion and open new avenues for exploring supersymmetric extensions of Galilean and Carrollian gravity theories.  

\end{abstract}
\end{titlepage}\newpage {} 

{\baselineskip=12pt \tableofcontents{}}

\section{Introduction}
Three-dimensional gravity continues to provide a valuable arena for exploring fundamental aspects of gravitational dynamics \cite{Banados:1992wn}, gauge formulations of gravity, and the interplay between geometry and symmetry. Among its most general realizations, the Mielke--Baekler (MB) model \cite{Mielke:1991nn} occupies a distinctive place, as it incorporates torsion in addition to curvature and admits a Chern--Simons (CS) formulation based on the so-called MB algebra \cite{Geiller:2020edh}. The presence of torsion makes the MB theory a natural starting point for investigating non-Riemannian phases of gravity, duality structures, and extended symmetry algebras. In particular, the MB model has been extensively studied in different contexts, including black hole solutions, holography, asymptotic symmetries, extended symmetries, and couplings to higher-spin gauge fields \cite{Blagojevic:2003uc,Blagojevic:2003vn,Blagojevic:2006hh,Giacomini:2006dr,Cvetkovic:2007sr,Klemm:2007yu,Blagojevic:2013bu,Peleteiro:2020ubv}. These developments highlight the versatility of the MB framework and motivate its extension to non-relativistic and supersymmetric regimes.

Recent developments in non-relativistic (NR) gravity, motivated by applications ranging from flat-space holography to condensed matter systems with Lifshitz or Galilean symmetries \cite{Son:2008ye,Balasubramanian:2008dm,Kachru:2008yh,Taylor:2008tg,Duval:2008jg,Bagchi:2009my,Hartnoll:2009sz,Bagchi:2009pe,Hoyos:2011ez,Son:2013rqa,Christensen:2013lma,Christensen:2013rfa,Abanov:2014ula,Hartong:2014oma,Hartong:2014pma,Hartong:2015wxa,Geracie:2015dea,Gromov:2015fda,Hartong:2015zia,Taylor:2015glc,Zaanen:2015oix,Devecioglu:2018apj}, have shown that NR limits of relativistic theories often require non-trivial extensions of the underlying symmetry algebras. 
In particular, consistent CS formulations of NR gravity demand not only finiteness in the NR limit but also the existence of non-degenerate invariant tensors, which in turn restrict the class of admissible NR algebras. In the case of the NR limit of the MB algebra, these difficulties have been overcome by introducing two central extensions, which arise from a NR contraction of a $U(1)$ enlargement of the MB algebra \cite{Concha:2023ejs}. 

Remarkably, as in its relativistic counterpart, NR MB gravity models depend on the parameters $(p,q)$, which allow one to switch on the spatial components of both curvature and torsion. Although non-vanishing torsion in the NR regime was first introduced in the Newton--Cartan framework by gauging the Schrödinger algebra \cite{Bergshoeff:2014uea}, the MB model not only provides a mechanism to incorporate torsion through a gauge-invariant action principle, but also unifies different NR gravity models within a single framework. In particular, for specific values of $(p,q)$, the MB theory reproduces the extended Newton--Hooke \cite{Papageorgiou:2010ud,Duval:2011mi,Hartong:2016yrf,Duval:2016tzi}, extended Bargmann \cite{Papageorgiou:2009zc,Bergshoeff:2016lwr}, and NR torsional gravity theories \cite{Concha:2021llq}. Non-vanishing torsion in NR settings has also been observed in the context of Lifshitz holography \cite{Christensen:2013lma} and the quantum Hall effect \cite{Geracie:2015dea}, further highlighting its physical relevance. Despite these developments, a consistent supersymmetric extension of NR MB gravity remains an open problem, which we address in this work.

In the presence of supersymmetry, the construction of a consistent NR supergravity theory becomes more involved, as the closure of the fermionic sector imposes additional constraints \cite{Bergshoeff:2022iyb}. It is then natural to ask whether the full MB supergravity theory admits a NR counterpart that preserves its gauge-theoretic and torsional structure. However, NR limits of supersymmetric CS theories are subtle: the contraction must simultaneously preserve the algebraic closure, the non-degeneracy of the invariant tensor, and the coupling between torsion and the fermionic sector. As we show, the minimal $\mathcal{N}=1$ MB superalgebra, although it admits a well-defined contraction, does not lead to a suitable NR superalgebra for defining a CS supergravity theory. 
This is due to the fact that the temporal translation generator (Hamiltonian operator) cannot be expressed as the anticommutator of two supercharges. This obstruction motivates the consideration of extended supersymmetry, which provides the appropriate framework for constructing a consistent NR MB supergravity theory.

A consistent NR limit can be achieved only by starting from an $\mathcal{N}=2$ MB superalgebra. However, to the best of our knowledge, there is no clear prescription for performing a NR contraction of the $\mathcal{N}=2$ superalgebra without introducing degeneracies. 
One way to overcome this issue is provided by the expansion mechanism \cite{Hatsuda:2001pp,deAzcarraga:2002xi,Izaurieta:2006zz,deAzcarraga:2007et}. In particular, the S-expansion method \cite{Izaurieta:2006zz} has proven to be a powerful tool for deriving NR superalgebras that admit non-degenerate invariant bilinear traces \cite{Concha:2024dap}, and hence for constructing well-defined CS supergravity actions \cite{deAzcarraga:2019mdn,Ozdemir:2019tby,Concha:2019mxx,Concha:2020tqx,Concha:2021jos,Concha:2024vql}.

In this work, by applying the S-expansion procedure to a $\mathfrak{so}(2)$ extension of the $\mathcal{N}=2$ MB superalgebra, we construct a new NR MB superalgebra that incorporates torsion through central extensions and admits a non-degenerate invariant tensor. This algebra contains, as particular cases, several known NR superalgebras, such as the extended Bargmann \cite{Bergshoeff:2016lwr,deAzcarraga:2019mdn,Concha:2020eam}, NR teleparallel \cite{Concha:2021llq}, and extended Newton--Hooke superalgebras \cite{Ozdemir:2019tby,Concha:2024dap}, while providing a unified framework that includes torsional and fermionic structures. We then construct the most general NR MB CS supergravity action, identifying the role played by torsion in both the bosonic and supersymmetric sectors. Our results provide the first systematic derivation of a torsional NR supergravity theory obtained from a fully relativistic CS supergravity with torsion. They clarify the interplay between supersymmetry, torsional extensions, and NR expansions, and establish a unified framework for non-relativistic supergravity theories in three dimensions.

The structure of the paper is as follows. Section~\ref{sec2} provides a concise review of three-dimensional MB gravity in its CS formulation. Section~\ref{sec3} develops the minimal supersymmetric extension of MB CS gravity. In Section~\ref{sec4}, we construct the NR counterpart of MB supergravity. We first derive the NR $\mathcal{N}=1$ MB superalgebra via an Inönü--Wigner contraction. Then, in order to obtain a well-defined NR supergravity theory, we consider a NR expansion of the $\mathfrak{so}(2)$ extension of the $\mathcal{N}=2$ MB superalgebra. 
The resulting NR MB supergravity action provides the most general supersymmetric extension of NR MB gravity and includes, as particular cases, various known NR supergravity actions for specific values of the $(p,q)$ parameters. Section~\ref{sec5} concludes the paper with a brief summary and a discussion of possible future directions. Additional details on the explicit (anti-)commutation relations of the $\mathcal{N}$-extended MB superalgebra are provided in Appendix~\ref{appa}. 

\section{Mielke-Baekler gravity theory à la Chern-Simons}\label{sec2}
The MB gravity model, originally introduced in \cite{Mielke:1991nn}, describes a three-dimensional gravity theory where both curvature and torsion are intrinsically non-vanishing. In contrast to standard Einstein gravity, torsion is not constrained to vanish and plays a dynamical role on equal footing with curvature. Subsequently, A CS formulation of MB gravity was developed in \cite{Geiller:2020edh, Caroca:2021njq} in terms of the gauge connection one-form ,
\begin{equation}
    A=W^{A}J_{A}+E^{A}P_{A}\,, \label{GC}
\end{equation}
where $W^{A}$ and $E^{A}$ denote the spin-connection and the dreibein, respectively. The generators $J_{A}$ and $P_{A}$ span the so-called MB algebra whose non-vanishing commutation relations read
\begin{eqnarray}
    \left[J_{A},J_{B}\right]&=&\epsilon_{ABC} J^{C}\,,\notag\\
    \left[J_{A},P_{B}\right]&=&\epsilon_{ABC} P^{C}\,,\notag\\
    \left[P_{A},P_{B}\right]&=&\epsilon_{ABC} \left(p\,J^{C}+q\,P^{C}\right)\,.\label{MB}
\end{eqnarray}
Here the indices $A,B=0,1,2$ are raised and lowered with the Minkowski metric $\eta_{AB}$ with the mostly plus signature convention. $\epsilon_{ABC}$ is the three-dimensional Levi Civita tensor which satisfies $\epsilon_{012}=-\epsilon^{012}=1$. On the other hand, $\left(p,q\right)$ are arbitrary constants which can be fixed to recover known relativistic Lie algebras (see Table~\ref{Table1}).
\begin{table}[!h]
\centering
    \begin{tabular}{|c|c|c|}
\hline
   & $p $ & $q$ \\ \hline\hline
Teleparallel algebra & 
   $0$ & $-2/\ell$ \\
   $\mathfrak{so}\left(2,2\right)$ algebra &
   $1/\ell^{2}$ & $0$ \\ 
   $\mathfrak{iso}\left(2,1\right)$ algebra & $0$ & $0$\\  \hline
\end{tabular}\captionsetup{font=footnotesize}\notag
\caption{Relativistic Lie algebras for different values of $p$ and $q$ in the MB algebra.}
\label{Table1}
\end{table}

The MB algebra admits the following non-vanishing components of the invariant tensor \cite{Geiller:2020edh}:
\begin{align}
    \langle J_{A}J_{B} \rangle &=\sigma_{2} \eta_{AB}\,, & \langle J_{A}P_{B} \rangle &=\sigma_{1} \eta_{AB}\,, & \langle P_{A}P_{B} \rangle &=\left(p\sigma_{2}+q\sigma_{1}\right) \eta_{AB}\,, \label{IT1}
\end{align}
where $\sigma_{1}$ and $\sigma_{2}$ are arbitrary constants which can be fixed to reproduce the original MB action \cite{Mielke:1991nn}. Indeed, a CS action for the MB algebra \eqref{MB} is obtained by considering the gauge connection one-form \eqref{GC} and the invariant tensor \eqref{IT1} into the general expression for the CS action:
\begin{eqnarray}
    I_{\text{CS}}&=& \int \left\langle AdA+\frac{2}{3}A^{3}\right\rangle \,.
\label{CS}
\end{eqnarray}
Then, the CS MB gravity action reads
\begin{eqnarray}
    I_{\text{CS}}[A]&=&\int\left(p\sigma_{1}+q\sigma_{3}\right)L_{0}[E^{A}]+\sigma_{1} L_{1}[E^{A},W^{A}]+\sigma_{2} L_{2}[W^{A}]+\sigma_{3} L_{3}[E^{A},W^{A}]\,,
\end{eqnarray}
where
\begin{eqnarray}
L_{0}[E^{A}] & = & \frac{1}{3} \epsilon_{ABC}E^{A}E^{B}E^{C}\,,\notag\\
L_{1} [E^{A},W^{A}]& = &2 E_{A}R^{A}\,,\notag\\
L_{2} [W^{A}]& = &  W^A d W_A + \frac{1}{3}\epsilon^{ABC}W_A W_B W_C\,,\notag\\
L_{3} [E^{A},W^{A}]& = & E_{A}T^{A}\,,\label{MBterms}
\end{eqnarray}
and
\begin{equation}
      \sigma_{3}=p \sigma_{2}+q \sigma_{1}\,. \label{sigma3}
\end{equation}
Here
\begin{align}
  R^{A}&=d W^{A}+\frac{1}{2}\epsilon^{ABC}W_{B}W_{C}\,,\notag\\
  T^{A}&=dE^{A}+\epsilon^{ABC}W_{B}E_{C}\,,
\end{align}
correspond to the Lorentz curvature and the torsion two-form, respectively. Thus, the MB CS action \eqref{MB} is described by a cosmological constant term $L_{0}[E^{A}]$, the Einstein-Hilbert term $L_{1}[E^{a},W^{A}]$, the exotic Lagrangian $L_{2}[W^{A}]$ and a torsional term $L_{3}[E^{A},W^{A}]$. Interestingly, the teleparallel CS gravity action \cite{Caroca:2021njq}, the AdS CS one \cite{Witten:1988hc,Zanelli:2005sa} and its Poincaré limit are reproduced by fixing the constants $p$ and $q$ as in Table \ref{Table1}. On the other hand, the original MB gravity model \cite{Mielke:1991nn},
\begin{eqnarray}
    I_{\text{MB}}[E^{A},W^{A}]=\int\sigma_{0}L_{0}[E^{A}]+\sigma_{1} L_{1}[E^{A},W^{A}]+\sigma_{2} L_{2}[W^{A}]+\sigma_{3} L_{3}[E^{A},W^{A}]\,,
\end{eqnarray}
is recovered after imposing
\begin{eqnarray}
    \sigma_{0}&=&p\sigma_{1}+q\sigma_{3}\,. \label{sigma0}
\end{eqnarray}
Let us note that the non-degeneracy of the invariant bilinear trace \eqref{IT1} ensures that the field equations are given by the vanishing of the curvature two-forms,
\begin{eqnarray}
 \mathcal{R}^{A}\left(W\right)&=&R^{A}+\frac{p}{2}\epsilon^{ABC}E_{B}E_{C}\,,\\
 \mathcal{R}^{A}\left(E\right)&=&T^{A}+\frac{q}{2}\epsilon^{ABC}E_{B}E_{C}\,.\label{RR}
\end{eqnarray}
which reproduces the MB field configuration given by a constant Lorentz curvature and constant torsion. For $q=0$, we obtain a vanishing torsion gravity model that coincides with the AdS CS model for $p=1/\ell^{2}=-\Lambda$ where $\ell$ is related to the cosmological constant $\Lambda$ and corresponds to the AdS radius. The cosmological constant appears as a source of torsion for $p=0$. In particular, imposing $q=-2/\ell$ reproduces the Teleparallel gravity field equations \cite{Geiller:2020edh, Caroca:2021njq}. Finally, the trivial case of $p=0$ and $q=0$ reproduces the Poincaré field configuration.

The CS formulation of MB gravity has proven to be a powerful framework to explore both the NR and ultra-relativistic (UR) regimes of three-dimensional gravity \cite{Concha:2023ejs,Concha:2025vhd}. In particular, the UR limit of the CS MB model allows for the introduction of non-vanishing torsion within three-dimensional Carrollian gravity. While supersymmetric extensions of the MB model have been previously investigated in \cite{Giacomini:2006dr}, the construction of a consistent NR limit at the supersymmetric level remains unexplored and, as we shall see, turns out to be more subtle. In this work, we address this issue by employing different strategies.

\section{On the supersymmetric extension of the Mielke-Baekler gravity}\label{sec3}
In this section, we construct the minimal supersymmetric extension of three-dimensional MB gravity. To this end, we consider a supersymmetric extension of the MB algebra \eqref{MB}, spanned by the Lorentz generators $J_A$, translational generators $P_A$, and fermionic charges $Q_\alpha$. These generators satisfy the following non-vanishing (anti)commutators,
\begin{eqnarray}
    \left[J_{A},J_{B}\right]&=&\epsilon_{ABC} J^{C}\,,\notag\\
    \left[J_{A},P_{B}\right]&=&\epsilon_{ABC} P^{C}\,,\notag\\
    \left[P_{A},P_{B}\right]&=&\epsilon_{ABC} \left(pJ^{C}+qP^{C}\right)\,,\notag\ \\
\left[J_{A},Q_{\alpha}\right]&=&-\frac{1}{2}(\gamma_A)_{\alpha}^{\,\beta}Q_\beta\,,\notag\\
\left[P_{A},Q_{\alpha}\right]&=&-\frac{\varepsilon_{+}}{2}(\gamma_A)_{\alpha}^{\,\beta}Q_\beta\,,\notag\\
\{Q_{\alpha},Q_{\beta}\}&=&-\left(\gamma^{A}C\right)_{\alpha\beta}\left(\varepsilon_{-}J_A+P_A\right)\,,\label{N1MB}
\end{eqnarray}
where $\alpha=1,2$ are spinorial indices, $\gamma^{a}$ are Dirac matrices in three dimensions, and $C$ is the charge conjugation matrix satisfying $C^{T}=-C$ and $C\gamma^{A}=(C\gamma^{A})^{T}$. For notational simplicity, we have also defined $\varepsilon_{\pm}$ as 
\begin{equation}
    \varepsilon_{\pm}=\sqrt{p+\frac{q^2}{4}}\pm\frac{q}{2}\,.\label{rmasmen}
\end{equation}
As in the relativistic bosonic case discussed in the previous section, specific choices of the constants 
$\left(p,q\right)$ allow one to recover known Lie superalgebras. Indeed, the Poincaré superalgebra, the teleparallel superalgebra presented in \cite{Caroca:2021njq} and the minimal supersymmetric extension of AdS are derived when the parameters are fixed as shown in Table~\ref{Table2}. 
\begin{table}[H]
\centering
    \begin{tabular}{|c|c|c|c|c|}
\hline
   & $p $ & $q$ & $\varepsilon_{+}$ & $\varepsilon_{-}$\\ \hline\hline
Teleparallel superalgebra & 
   $0$ & $-2/\ell$ & $0$ & $2/\ell$ \\ 
 AdS superalgebra &
   $1/\ell^{2}$ & $0$ & $1/\ell$ & $1/\ell$\\ 
   Poincaré superalgebra & $0$ & $0$& $0$ & $0$\\  \hline
\end{tabular}\captionsetup{font=footnotesize}\notag
\caption{Superalgebras for different values of $p$ and $q$ in the $\mathcal{N}=1$ MB superalgebra.}
\label{Table2}
\end{table}

The gauge connection one-form $A$ for the MB superalgebra \eqref{N1MB} is 
\begin{equation}
    A= W^{A}J_{A}+E^{A} P_{A}+\bar{\Psi}\,Q\,,\label{one-formN1MB}
\end{equation}
where $W^A$ denotes the spin connection one-form, $E^A$ represents the dreibein, and $\Psi$ is a Majorana fermionic field corresponding to the gravitino. The associated curvature two-form is given by
\begin{equation}
    F= \hat{\mathcal{R}}^{A} J_{A}+\mathcal{T}^{A} P_{A}+\nabla\bar{\Psi}\, Q\,,\label{two-formN1MB}
\end{equation}
where
\begin{eqnarray}
\hat{\mathcal{R}}^{A}&=&\mathcal{R}^{A}(W)+\frac{\varepsilon_{-}}{2}\bar{\Psi}\gamma^{A}\Psi \,,\notag\\
 \mathcal{T}^{A}&=&\mathcal{R}^{A}(E)+\frac{1}{2}\bar{\Psi}\gamma^{A}\Psi \,,\notag\\
 \nabla\Psi&=&d\Psi+\frac{1}{2}W^{A}\gamma_A\Psi+\frac{\varepsilon_{+}}{2}E^{A}\gamma_A\Psi\,.\label{curvaN1MB}
\end{eqnarray}
Here $\mathcal{R}^{A}(W)$ and $\mathcal{R}^{A}(E)$ are defined in \eqref{RR}. $\mathcal{T}^{A}$ is a super-torsion, and $\nabla\Psi$ defines the covariant derivative of the gravitino. The MB superalgebra \eqref{N1MB} admits a non-degenerate invariant tensor, whose non-vanishing components are given by \eqref{IT1} along with
\begin{eqnarray}
    \langle Q_{\alpha},Q_{\beta}\rangle &=& 2\left(\varepsilon_{-}\sigma_2+\sigma_1\right)C_{\alpha\beta}\,, \label{invtensor2}
\end{eqnarray}
Then, considering the gauge connection one-form \eqref{one-formN1MB} along with the non-vanishing components of the invariant tensor given in \eqref{IT1} and \eqref{invtensor2}, in the CS action \eqref{CS}, we get
\begin{eqnarray}
    I_{\text{sMB}}[E^{A},W^{A},\Psi]=\int\sigma_{0}L_{0}[E^{A}]+\sigma_{1} L_{1}[E^{A},W^{A},\Psi]+\sigma_{2} L_{2}[W^{A},\Psi]+\sigma_{3} L_{3}[E^{A},W^{A}]\,,
\end{eqnarray}
where $\sigma_0$ and $\sigma_3$ are related to $\sigma_1$ and $\sigma_2$ as in eqs. \eqref{sigma3} and \eqref{sigma0}. Here, the fermionic contributions along $\sigma_1$ and $\sigma_2$ are explicitly given by
\begin{eqnarray}
L_{1} [E^{A},W^{A},\Psi]& = &2 E_{A}R^{A}-2\bar{\Psi}\nabla\Psi\,,\notag\\
L_{2} [W^{A},\Psi]& = &  W^A d W_A + \frac{1}{3}\epsilon^{ABC}W_A W_B W_C-2\varepsilon_{-}\bar{\Psi}\nabla\Psi\,.\label{MBfermterms}
\end{eqnarray}

The CS action presented above corresponds to the minimal supersymmetric extension of the MB gravity theory. This action was originally constructed in \cite{Giacomini:2006dr}, where a duality map between different gravity theories was proposed to obtain the supersymmetric extension of the MB model. In the present work, we have shown that the same supergravity theory admits an equivalent formulation within the CS framework, based on the MB superalgebra \eqref{N1MB}.

\section{Non-relativistic regime of the Mielke-Baekler supergravity}\label{sec4}

In this section, we derive the NR version of the MB supergravity theory. Our results correspond to a supersymmetric extension of the NR MB gravity model analyzed in \cite{Concha:2023ejs} and briefly discussed in Section 2. We show that, in order to obtain a consistent NR supergravity theory, it is necessary to start from the $\mathcal{N}=2$ MB superalgebra. 

By applying the S-expansion method \cite{Izaurieta:2006zz}, and inspired by the results obtained in \cite{Concha:2024dap}, we construct the corresponding NR MB superalgebra together with its non-degenerate invariant tensor, which constitutes the essential ingredient for defining a well-defined CS supergravity action. Before presenting this construction, we first show that a well-defined NR limit of the minimal supersymmetric extension of the MB algebra given in \eqref{N1MB} can indeed be performed; however, it does not yield a suitable superalgebra for the construction of a consistent NR supergravity theory.

\subsection{Non-relativistic $\mathcal{N}=1$ Mielke-Baekler superalgebra}
Following the approach of \cite{Concha:2023ejs}, the NR $\mathcal{N}=1$ MB superalgebra can be derived from its relativistic counterpart \eqref{N1MB} by means of an Inönü–Wigner contraction. To perform this contraction, we first decompose the Lorentz index as $A=(0,a)$ with $a=1,2$, thereby distinguishing temporal and spatial components. This decomposition naturally leads to the following splitting of the relativistic generators:
\begin{align}
    J_A\rightarrow \{J_0\equiv J,J_a\}\,,\qquad P_A\rightarrow \{P_0\equiv H,P_a\}\,.
\end{align}
Thus, the MB superalgebra \eqref{N1MB} is defined by the following non-vanishing commutation relations,
\begin{align}
    \left[J_{a},J_{b}\right]&=-\epsilon_{ab} J\,,&
    \left[J,J_{a}\right]&=\epsilon_{ab} J_{b}\,,\notag\\
    \left[J_{a},P_{b}\right]&=-\epsilon_{ab} H\,,&
    \left[J,P_{a}\right]&=\epsilon_{ab} P_{b}\,,\notag\\
    \left[H,J_{a}\right]&=\epsilon_{ab} P_{b}\,,&
    \left[P_{a},P_{b}\right]&=\epsilon_{ab} \left(pJ+qH\right)\,,\notag\ \\
    \left[H,P_{a}\right]&=\epsilon_{ab} \left(pJ_{b}+qP_{b}\right)\,,&
    \left[J_{a},Q_{\alpha}\right]&=-\frac{1}{2}(\gamma_a)_{\alpha}^{\,\beta}Q_\beta\,,\notag\\
    \left[J_{0},Q_{\alpha}\right]&=-\frac{1}{2}(\gamma_0)_{\alpha}^{\,\beta}Q_\beta\,,&
    \left[P_{a},Q_{\alpha}\right]&=-\frac{\varepsilon_{+}}{2}(\gamma_a)_{\alpha}^{\,\beta}Q_\beta\,,\notag\\
    \left[H,Q_{\alpha}\right]&=-\frac{\varepsilon_{+}}{2}(\gamma_0)_{\alpha}^{\,\beta}Q_\beta\,, \label{N1MBdescompose1}
\end{align}
together with the anticommutator
\begin{align}
    \{Q_{\alpha},Q_{\beta}\}&=-\left(\gamma^{0}C\right)_{\alpha\beta}\left(\varepsilon_{-}J+H\right) -\left(\gamma^{a}C\right)_{\alpha\beta}\left(\varepsilon_{-}J_a+P_a\right)\,.  \label{N1MBdescompose2}
\end{align}
Now, under the rescaling
\begin{align}
    J &\rightarrow J\,, & J_{a} &\rightarrow \xi G_{a}\,, & H &\rightarrow H\,, & P_{a} &\rightarrow  \xi P_{a}& Q_{\alpha} \rightarrow \xi^{1/2} Q_{\alpha}\,,
\end{align}
it is straightforward to verify that, in the $\xi \to \infty$ limit, the algebra \eqref{N1MBdescompose1}-\eqref{N1MBdescompose2} reduces to the NR $\mathcal{N}=1$ MB superalgebra, whose generators satisfy the following (anti)commutation relations
\begin{align}
      \left[J,G_{a}\right]&=\epsilon_{ab} G_{b}\,,&
    \left[J,P_{a}\right]&=\epsilon_{ab} P_{b}\,,\notag\\
     \left[H,G_{a}\right]&=\epsilon_{ab} P_{b}\,,&
    \left[H,P_{a}\right]&=\epsilon_{ab} \left(pG_{b}+qP_{b}\right)\,,\notag\ \\
\left[J,Q_{\alpha}\right]&=-\frac{1}{2}(\gamma_0)_{\alpha}^{\,\beta}Q_\beta\,,&
\left[H,Q_{\alpha}\right]&=-\frac{\varepsilon_{+}}{2}(\gamma_0)_{\alpha}^{\,\beta}Q_\beta\,,\notag\\
\{Q_{\alpha},Q_{\beta}\}&=-\left(\gamma^{a}C\right)_{\alpha\beta}\left(\varepsilon_{-}G_a+P_a\right)\,.
\end{align}
This algebra corresponds to the minimal supersymmetric extension of the torsional galilean-AdS algebra introduced in \cite{Matulich:2019cdo}, 
being a particular case (with $M=S=0$) of the NR MB algebra introduced in \cite{Concha:2023ejs}. It is worth noting that the anticommutator of two supercharges generates spatial translations but not temporal ones. In relativistic superalgebras, the closure of $\{Q,Q\}$ yields the full set of translation generators. In the NR case, however, a minimal consistency requirement is that the supercharges close on the temporal translation generator $H$, while in extended realizations they may also involve the spatial translations $P_a$. 

Since the contraction procedure fails to meet this requirement, it is natural to consider a more general framework that includes contractions as a particular case. In this context, one may employ the expansion method of Lie (super)algebras based on abelian semigroups, which has the additional advantage of systematically providing invariant tensors for the expanded algebra. Following \cite{Izaurieta:2006zz}, the standard contraction can be recovered using the semigroup $S_{E}^{(1)}$, while the first non-trivial expansion arises from the semigroup $S_{E}^{(2)}$. However, even in this case, the same obstruction persists, namely that the anticommutator $\{Q,Q\}$ does not generate the temporal translation $H$. These observations, together with previous results obtained in \cite{Concha:2024dap}, where extended supersymmetry was shown to be necessary to construct consistent NR supergravity theories with non-degenerate invariant tensors, motivate us to consider the $\mathcal{N}=2$ extension as the minimal consistent framework, rather than its technically simpler $\mathcal{N}=1$ counterpart.

\subsection{Non-relativistic $\mathcal{N}=2$ Mielke-Baekler superalgebra}
A non-relativistic MB superalgebra can be derived by applying the $S$-expansion method to the $\mathfrak{so}\left(2\right)$ extension of the $\mathcal{N}=2$ MB superalgebra (see Appendix \eqref{appa}), whose generators satisfy the following (anti)-commutation relations,
\begin{eqnarray}
    \left[J_{A},J_{B}\right]&=&\epsilon_{ABC} J^{C}\,,\notag\\
    \left[J_{A},P_{B}\right]&=&\epsilon_{ABC} P^{C}\,,\notag\\
    \left[P_{A},P_{B}\right]&=&\epsilon_{ABC} \left(pJ^{C}+qP^{C}\right)\,,\notag\ \\
\left[J_{A},Q_{\alpha}^{i}\right]&=&-\frac{1}{2}(\gamma_A)_{\alpha}^{\,\beta}Q_\beta^{i}\,,\notag\\
\left[P_{A},Q_{\alpha}^{i}\right]&=&-\frac{\varepsilon_{+}}{2}(\gamma_A)_{\alpha}^{\,\beta}Q_\beta^{i}\,,\notag\\
  \left[T,Q_{\alpha}^{i}\right]&=&  \frac{\epsilon^{ij}}{2}Q_{\alpha}^{j}\,,\notag \\
\left[Q_{\alpha}^{i},Q_{\beta}^{j}\right]&=&-\delta^{ij
}\left(\gamma^{A}C\right)_{\alpha\beta}\left(\varepsilon_{-}J_A+P_A\right)-C_{\alpha\beta}\epsilon^{ij}\left(U+\varepsilon_{-}T\right)\,,\label{N2MB}
\end{eqnarray}
where $Q^{i}_\alpha$ are Majorana spinor charges, and the constants $\varepsilon_{\pm}$ were defined in \eqref{rmasmen}. The presence of the bosonic generator $U$ and the $\mathfrak{so}\left(2\right)$ automorphism generator $T$ is required to ensure the non-degeneracy of the invariant bilinear trace in the Poincaré limit $\left(p,q,\varepsilon_\pm\right)\rightarrow 0$. 

As before, we decompose the index $A$ into temporal and spatial components,
\begin{equation}
    A\rightarrow(0,a)\,, \quad a=1,2\,.
\end{equation}
This leads to the following splitting of the relativistic generators:
\begin{align}
    J_A\rightarrow \{J_0\equiv J,J_a\}\,,\qquad P_A\rightarrow \{P_0\equiv H,P_a\}\,.
\end{align}
In parallel, we reorganize the fermionic generators as
\begin{equation}
    Q_\alpha^{\pm}=\frac{1}{\sqrt{2}}\left(Q_\alpha^1\pm (\gamma_0)_{\alpha}^{\,\,\beta}Q_\beta^2 \right)\,,
\end{equation}
such that the $\mathcal{N}=2$ MB superalgebra takes a form better suited for the NR limit. The non-vanishing commutation relations are given by
\begin{align}
   \left[J,J_{a}\right]&=\epsilon_{ab} J_{b}\,,&   \left[J,P_{a}\right]&=\epsilon_{ab} P_{b}\notag\\
    \left[J_{a},J_{b}\right]&=-\epsilon_{ab} J\,,\notag & \left[J_{a},P_{b}\right]&=-\epsilon_{ab} H\,,\notag\\
  \left[H,J_{a}\right]&=\epsilon_{ab} P_{b}\,,  &\left[P_{a},P_{b}\right]&=-\epsilon_{ab} \left(pJ+qH\right)\,,\notag \\
\left[H,P_{a}\right]&=\epsilon_{ab} \left(pJ_b+q P_b\right)\, &\left[T,Q_{\alpha}^{\pm}\right]&=\pm\frac{1}{2}(\gamma_0)_{\alpha}^{\,\beta}Q_\beta^{\pm}\,,\notag\\
    \left[J,Q_{\alpha}^{\pm}\right]&=-\frac{1}{2}(\gamma_0)_{\alpha}^{\,\beta}Q_\beta^{\pm}\,,&   \left[H,Q_{\alpha}^{\pm}\right]&=-\frac{\varepsilon_{+}}{2}(\gamma_0)_{\alpha}^{\,\beta}Q_\beta^{\pm}\,,\notag\\
    \left[J_a,Q_{\alpha}^{\pm}\right]&=-\frac{1}{2}(\gamma_a)_{\alpha}^{\,\beta}Q_\beta^{\mp}\,,&   \left[P_a,Q_{\alpha}^{\pm}\right]&=-\frac{\varepsilon_{+}}{2}(\gamma_a)_{\alpha}^{\,\beta}Q_\beta^{\mp}\,,
\end{align}
while the non-vanishing anticommutators read
\begin{align}
    \{Q_{\alpha}^{-},Q_{\beta}^{-}\}&=-\left(\gamma^{0}C\right)_{\alpha\beta}\left(\varepsilon_{-}J+H\right)+\left(\gamma^{0}C\right)_{\alpha\beta}\left(U+\varepsilon_{-}T\right)\,,\notag \\
    \{Q_{\alpha}^{+},Q_{\beta}^{+}\}&=-\left(\gamma^{0}C\right)_{\alpha\beta}\left(\varepsilon_{-}J+H\right)-\left(\gamma^{0}C\right)_{\alpha\beta}\left(U+\varepsilon_{-}T\right)\,,\notag \\
   \{Q_{\alpha}^{+},Q_{\beta}^{-}\}&=-\left(\gamma^{a}C\right)_{\alpha\beta}\left(\varepsilon_{-}J_a+P_a\right)\,.
\end{align}
A natural next step would be to attempt a contraction of this algebra or, equivalently, as discussed in the previous section, an expansion using the semigroup $S_{E}^{(1)}$, based on the decomposition given in Eq.~(\ref{desc-N2}). However, the algebra obtained through this procedure admits a degenerate invariant tensor, which prevents the construction of a consistent CS supergravity action. This limitation motivates the consideration of a non-trivial expansion of the relativistic superalgebra, capable of yielding a non-degenerate invariant structure. We then perform the expansion of this relativistic superalgebra following the procedure employed in \cite{Concha:2023bly,Concha:2024dap}. For this purpose, we consider the semigroup $S_{E}^{(2)}=\{\lambda_0,\lambda_1,\lambda_2,\lambda_3\}$, whose elements satisfy the following multiplication law,
\begin{equation}
\begin{tabular}{l|llll}
$\lambda _{3}$ & $\lambda _{3}$ & $\lambda _{3}$ & $\lambda _{3}$ & $\lambda
_{3}$ \\
$\lambda _{2}$ & $\lambda _{2}$ & $\lambda _{3}$ & $\lambda _{3}$ & $\lambda
_{3}$ \\
$\lambda _{1}$ & $\lambda _{1}$ & $\lambda _{2}$ & $\lambda _{3}$ & $\lambda
_{3}$ \\
$\lambda _{0}$ & $\lambda _{0}$ & $\lambda _{1}$ & $\lambda _{2}$ & $\lambda
_{3}$ \\ \hline
& $\lambda _{0}$ & $\lambda _{1}$ & $\lambda _{2}$ & $\lambda _{3}$%
\end{tabular}
\label{sMBsemi}
\end{equation}
Here, $\lambda_3=0_s$ denotes the zero element of the semigroup, satisfying $0_s\lambda_k=0_s$, $k = 1,2,3$. Before carrying out the S-expansion procedure, we introduce a specific subspace decomposition of the 
$\mathcal{N}=2$ MB superalgebra as follows,
\begin{equation}
    V_0 = \lbrace J, H, T, U,Q^+_\alpha \rbrace\,, \qquad V_1 = \lbrace J_a, P_a, Q^-_\alpha \rbrace\,, \label{desc-N2}
\end{equation}
which satisfies 
\begin{eqnarray}
[V_0,V_0]\subset V_0\,,\qquad [V_0,V_1]\subset V_1\,, \qquad [V_1,V_1]\subset V_0\,.
\end{eqnarray}
Similarly, we now consider the decomposition of the relevant semigroup as $S_{E}^{(2)}=S_{0} \cup S_{1}$, where
\begin{eqnarray}
S_0&=&\{\lambda_0,\lambda_2,\lambda_3\}\,, \notag \\
S_1&=&\{\lambda_1,\lambda_3\} \,. 
\label{decomp}
\end{eqnarray}
This decomposition is said to be resonant, as it satisfies the same structure as the subspace decomposition of the algebra. In particular, it fulfills
\begin{equation}
    S_0\cdot S_0\subset S_0\,,\qquad \ S_0\cdot S_1\subset S_1\,,\qquad \ S_1\cdot S_1\subset S_0\,.\label{semidecomp}
\end{equation}
Following \cite{Izaurieta:2006zz}, we extract a resonant subalgebra from the $S_{E}^{(2)}$-expansion of the $\mathcal{N}=2$ MB superalgebra,
\begin{eqnarray}
    \mathfrak{G}=\left(S_0\times V_0\right)\oplus \left(S_1\times V_1\right)\,,
\end{eqnarray}
and subsequently perform a $0_s$-reduction by imposing $\lambda_3 T_{A}=0$. As a result, we obtain a new NR expanded superalgebra spanned by the generators
\begin{equation}
\{{J},G_{a},{S},{H},{P}_{a},{M},{T}_1,{T}_2,{U}_1,{U}_2,{Q}_{\alpha }^{+},{R}_{\alpha},{Q}_{\alpha
}^{-}\}\,.
\end{equation}
The above procedure can be represented schematically as in Table~\ref{Sexpa}.

\renewcommand{\arraystretch}{1.4}
\begin{table}[H]
\centering
    \begin{tabular}{lll}
\multicolumn{1}{l|}{$\lambda_3$} & \multicolumn{1}{|l}{\cellcolor[gray]{0.8}} & \multicolumn{1}{|l|}{\cellcolor[gray]{0.8}} \\ \hline
\multicolumn{1}{l|}{$\lambda_2$} & \multicolumn{1}{|l}{${S},\ \,{M},\ \,{T}_2,\ \,{U}_2,\ \,{R}_{\alpha}$} & \multicolumn{1}{|l|}{\cellcolor[gray]{0.8}} \\ \hline
\multicolumn{1}{l|}{$\lambda_1$} & \multicolumn{1}{|l}{\cellcolor[gray]{0.8}} & \multicolumn{1}{|l|}{${G}_a,\ \,{P}_a,\ \,{Q}^{-}_{\alpha}$} \\ \hline
\multicolumn{1}{l|}{$\lambda_0$} & \multicolumn{1}{|l}{$ {J},\ \ {H},\ \ {T}_1,\ \,{U}_1,\ \,{Q}^{+}_{\alpha}$} & \multicolumn{1}{|l|}{\cellcolor[gray]{0.8}} \\ \hline
\multicolumn{1}{l|}{} & \multicolumn{1}{|l}{$J,\ \ H,\ \ {T},\ \ \,U,\ \ \,Q^{+}_{\alpha}$} & \multicolumn{1}{|l|}{$J_{a},\ \ P_{a},\ \ Q^{-}_{\alpha}$} 
\end{tabular}
\captionsetup{font=footnotesize}
\caption{NR expanded generators expressed in terms of the $\mathcal{N}=2$ super MB generators through the $S_E^{(2)}$ semigroup elements.}
\label{Sexpa}
\end{table}
The NR generators satisfy the following non-vanishing commutation relations:
\begin{align}
\left[ J,G_{a}\right] &=\epsilon _{ab}{G}_{b}\,, &
\left[ J,P_{a}\right] &=\epsilon _{ab}{P}_{b}\,, &
\left[G_{a},{P}_{b}\right]&=-\epsilon_{ab}M\,,  \notag
\\
\left[ G_{a},G_{b}\right] &=-\epsilon _{ab}S\,,&
\left[ P_{a},P_{b}\right] &=-\epsilon _{ab}\left(p S + q M\right)\,,&
\left[ H,G_{a}\right]& =\epsilon _{ab}P_{b}\,,  \notag \\
\left[ H,P_{a}\right]& =\epsilon _{ab}\left(pG_{b}+qP_{b}\right)\,, &
     \left[J,Q_{\alpha}^{\pm}\right]&=-\frac{1}{2}(\gamma_0)_{\alpha}^{\,\beta}Q_\beta^{\pm}\,,&  \left[J,R_{\alpha}\right]&=-\frac{1}{2}(\gamma_0)_{\alpha}^{\,\beta}R_\beta\,, \notag \\   \left[S,Q_{\alpha}^{+}\right]&=-\frac{1}{2}(\gamma_0)_{\alpha}^{\,\beta}R_\beta\,, &
     \left[H,Q_{\alpha}^{\pm}\right]&=-\frac{\varepsilon_{+}}{2}(\gamma_0)_{\alpha}^{\,\beta}Q_\beta^{\pm}\,,& \left[H,R_{\alpha}\right]&=-\frac{\varepsilon_{+}}{2}(\gamma_0)_{\alpha}^{\,\beta}R_\beta\,, \notag \\   \left[M,Q_{\alpha}^{+}\right]&=-\frac{\varepsilon_{+}}{2}(\gamma_0)_{\alpha}^{\,\beta}R_\beta\,, &
     \left[G_a,Q_{\alpha}^{+}\right]&=-\frac{1}{2}(\gamma_a)_{\alpha}^{\,\beta}Q_\beta^{-}\,,& \left[G_a,Q_{\alpha}^{-}\right]&=-\frac{1}{2}(\gamma_a)_{\alpha}^{\,\beta}R_\beta\,,\notag \\\left[P_a,Q_{\alpha}^{+}\right]&=-\frac{\varepsilon_{+}}{2}(\gamma_a)_{\alpha}^{\,\beta}Q_\beta^{-}\,, &
     \left[P_a,Q_{\alpha}^{-}\right]&=-\frac{\varepsilon_{+}}{2}(\gamma_a)_{\alpha}^{\,\beta}R_\beta\,,&  \left[T_1,Q_{\alpha}^{\pm}\right]&=\pm\frac{1}{2}(\gamma_0)_{\alpha}^{\,\beta}Q_\beta^{\pm}\,,\notag \\ \left[T_1,R_{\alpha}\right]&=\frac{1}{2}(\gamma_0)_{\alpha}^{\,\beta}R_\beta\,, &
     \left[T_2,Q_{\alpha}^{+}\right]&=\frac{1}{2}(\gamma_0)_{\alpha}^{\,\beta}R_\beta\,,\label{sMBNR1}
     \end{align}
together with the following non-vanishing anitcommutators:
     \begin{align}
     \{Q_{\alpha}^{-},Q_{\beta}^{-}\}&=-\left(\gamma^{0}C\right)_{\alpha\beta}\left(\varepsilon_{-}S+M\right)+\left(\gamma^{0}C\right)_{\alpha\beta}\left(U_2+\varepsilon_{-}T_2\right)\,,\notag \\
     \{Q_{\alpha}^{+},Q_{\beta}^{+}\}&=-\left(\gamma^{0}C\right)_{\alpha\beta}\left(\varepsilon_{-}J+H\right)-\left(\gamma^{0}C\right)_{\alpha\beta}\left(U_1+\varepsilon_{-}T_1\right)\,,\notag \\
     \{Q_{\alpha}^{+},R_{\beta}\}&=-\left(\gamma^{0}C\right)_{\alpha\beta}\left(\varepsilon_{-}S+M\right)-\left(\gamma^{0}C\right)_{\alpha\beta}\left(U_2+\varepsilon_{-}T_2\right)\,,\notag \\
     \{Q_{\alpha}^{+},Q_{\beta}^{-}\}&=-\left(\gamma^{a}C\right)_{\alpha\beta}\left(\varepsilon_{-}G_a+P_a\right)\,. \label{sMBNR3}
\end{align}
The bosonic NR subalgebra spanned by $\{J,G_{a},H,P_{a},S,M\}$ corresponds to the NR MB algebra introduced in \cite{Concha:2023ejs}. It contains spatial translations $P_a$, spatial rotations $J$, Galilean boosts $G_a$, time translations $H$, and two central charges $S$ and $M$. Its supersymmetric extension requires three spinor charges $Q_{\alpha}^{+}$, $Q_{\alpha}^{-}$ and $R_{\alpha}$, in analogy with the extended Bargmann superalgebra \cite{Bergshoeff:2016lwr}. Moreover, additional bosonic generators $\{T_1,T_2\}$ and central charges $\{U_1,U_2\}$ are necessary to ensure the non-degeneracy of the invariant bilinear trace, even in the flat limit $\ell\rightarrow\infty$. The generators $\{T_1,T_2\}$ arise as the S-expansion of the R-symmetry generator $T$ of the $\mathfrak{so}\left(2\right)$ extension of the $\mathcal{N}=2$ MB superalgebra.

Similarly to the purely bosonic case, different known NR superalgebras can be obtained from \eqref{sMBNR1}--\eqref{sMBNR3} for particular values of the parameters $(p,q)$. In particular, the extended Bargmann superalgebra \cite{Bergshoeff:2016lwr,deAzcarraga:2019mdn,Concha:2020eam}, the torsional NR superalgebra constructed in \cite{Concha:2021llq}, and the supersymmetric extension of the extended Newton--Hooke algebra \cite{Ozdemir:2019tby,Concha:2024dap} arise for the choices of parameters summarized in Table~\ref{Table3}.
\begin{table}[!h]
\centering
    \begin{tabular}{|c|c|c|c|c|}
\hline
   & $p $ & $q$ & $\varepsilon_{+}$ & $\varepsilon_{-}$\\ \hline\hline
NR Teleparallel superalgebra & 
   $0$ & $-2/\ell$ & $0$ & $2/\ell$ \\ 
Extended Newton-Hooke superalgebra &
   $1/\ell^{2}$ & $0$ & $1/\ell$ & $1/\ell$\\ 
   Extended Bargmann superalgebra & $0$ & $0$& $0$ & $0$\\  \hline
\end{tabular}\captionsetup{font=footnotesize}\notag
\caption{NR Superalgebras for different values of $p$ and $q$ in the NR MB superalgebra.}
\label{Table3}
\end{table}

\subsection{Non-relativistic Chern-Simons Mielke-Baekler supergravity}
We now turn to the construction of the most general NR MB CS supergravity action, whose underlying symmetry is given by the NR MB superalgebra \eqref{sMBNR1}-\eqref{sMBNR3}. To this end, we introduce the gauge connection one-form,
\begin{equation}
    A=\omega J+\omega^a G_a+\tau H +e^{a} P_{a}+mM+sS+t_1 T_1+t_2 T_2+u_1U_1+u_2 U_2+\bar{\psi}^{+} Q^{+}+\bar{\psi}^{-} Q^{-}+\bar{\rho} R\,.\label{NRoneform}
\end{equation}
The corresponding curvature two-form is given by
\begin{eqnarray}
{F} &=&R(\omega ){J}%
+R^{a}(\omega ^{b}){G}_{a}+R(\tau ){H}+R^{a}(e^{b}){P}_{a}+R(m)
{M}+R(s){S} \notag \\
& & + R(t_1) {T}_1 + R(t_2){T}_2 + R(u_1) {U}_1 + R(u_2) {U}_2 + \nabla \bar{\psi}^+ {Q}^+ + \nabla \bar{\psi}^-{Q}^- + \nabla \bar{\rho} {R} \,,\label{supertorcuv}
\end{eqnarray}%
where the bosonic curvatures are given by
\begin{eqnarray}
R(\omega ) &=& d \omega + \frac{\varepsilon_{-}}{2} \bar{\psi}^+ \gamma^0 \psi^+ \,,  \notag \\
R^{a}(\omega ^{b}) &=&d\omega ^{a}+\epsilon ^{ac}\omega \omega _{c} + p\epsilon^{ac}\tau e_{c} + \varepsilon_{-} \bar{\psi}^+ \gamma^a \psi^- \,, \notag \\
R(\tau ) &=& d \tau + \frac{1}{2} \bar{\psi}^+ \gamma^0 \psi^+ \,,  \notag \\
R^{a}(e^{b}) &=& de^{a}+\epsilon ^{ac}\omega e_{c}+\epsilon ^{ac}\tau \omega
_{c}+q\epsilon^{ac}\tau e_{c} + \bar{\psi}^+ \gamma^a \psi^- \,,  \notag \\
R(m) &=& dm+\epsilon ^{ac}e_{a}\omega _{c}+\frac{q}{2}\epsilon^{ac}e_{a}e_{c} + \frac{1}{2} \bar{\psi}^- \gamma^0 \psi^- + \bar{\psi}^+ \gamma^0 \rho \,,  \notag \\
R(s) &=&ds+\frac{1}{2}\epsilon ^{ac}\omega _{a}\omega _{c} + \frac{p}{2}\epsilon^{ac}e_{a}e_{c} + \frac{\varepsilon_{-}}{2} \bar{\psi}^- \gamma^0 \psi^- + \varepsilon_{-} \bar{\psi}^+ \gamma^0 \rho \,, \notag \\
R(t_1) &=& dt_1 + \frac{\varepsilon_{-}}{2} \bar{\psi}^+ \gamma^0 \psi^+ \,, \notag \\
R(t_2) &=& dt_2 - \frac{\varepsilon_{-}}{2} \bar{\psi}^- \gamma^0 \psi^- + {\varepsilon_{-}} \bar{\psi}^+ \gamma^0 \rho \,, \notag \\
R(u_1) &=& du_1 + \frac{1}{2} \bar{\psi}^+ \gamma^0 \psi^+ \,, \notag \\
R(u_2) &=& du_2 - \frac{1}{2} \bar{\psi}^- \gamma^0 \psi^- + \bar{\psi}^+ \gamma^0 \rho \,, \label{N2curvNRone}
\end{eqnarray}
while the fermionic components are given by
\begin{eqnarray}
\nabla \psi^+ &=& d\psi^+ + \frac{1}{2} \omega \gamma_0 \psi^+ +\frac{\varepsilon_{+}}{2}\tau\gamma_0 \psi^+ + \frac{t_1}{2} \gamma_0 \psi^+ \,,  \notag \\
\nabla \psi^- &=& d\psi^- + \frac{1}{2} \omega \gamma_0 \psi^- +\frac{\varepsilon_{+}}{2}\tau\gamma_0 \psi^- + \frac{1}{2} \omega^a \gamma_a \psi^++\frac{\varepsilon_{+}}{2}e^a \gamma_a \psi^+ - \frac{t_1}{2} \gamma_0 \psi^- \,,  \notag \\
\nabla \rho &=& d\rho + \frac{1}{2} \omega \gamma_0 \rho+\frac{\varepsilon_{+}}{2}\tau\gamma_0 \rho + \frac{1}{2} s \gamma_0 \psi^+ + \frac{\varepsilon_{+}}{2} m \gamma_0 \psi^++ \frac{1}{2} \omega^a \gamma_a \psi^- +\frac{\varepsilon_{+}}{2} e^a \gamma_a \psi^-\notag \\
&&+\frac{t_2}{2} \gamma_0 \psi^+ + \frac{t_1}{2} \gamma_0 \rho \,.  \label{N2curvNRtwo}
\end{eqnarray}

The non-vanishing components of the invariant tensor for the NR MB superalgebra are given by those of the NR MB algebra \cite{Concha:2023ejs},
\begin{eqnarray}
\left\langle {J}{S}\right\rangle &=&-\tilde{\sigma}_2\,,  \notag \\
\left\langle {G}_{a}{G}_{b}\right\rangle &=&\tilde{\sigma}%
_{2}\delta _{ab}\,,  \notag \\
\left\langle {G}_{a}{P}_{b}\right\rangle &=&\tilde{\sigma}%
_{1}\delta _{ab}\,,  \notag \\ 
\left\langle {H}{S}\right\rangle &=&\left\langle {M}{%
J}\right\rangle =-\tilde{\sigma}_{1}\,,  \notag \\
\left\langle{P}_{a}{P}_{b}\right\rangle &=&\left(p\tilde{\sigma}_{2}+q \tilde{\sigma}_{1}\right)\delta_{ab}\,,\notag \\
\left\langle {H}{%
M}\right\rangle &=&-\left(p\tilde{\sigma}_{2}+q \tilde{\sigma}_{1}\right)\,,  \label{sMBIT1}
\end{eqnarray}
together with
\begin{eqnarray}
    \left\langle T_1 T_2 \right\rangle &=& \tilde{\sigma}_2\,, \notag\\
\left\langle T_1 U_2 \right\rangle &=& \left\langle T_2 U_1 \right\rangle = \tilde{\sigma}_1-q\tilde{\sigma}_2\,, \notag\\
\left\langle U_1 U_2 \right\rangle &=& \left(\frac{q}{2}-\varepsilon_{-}\right)\tilde{\sigma}_1+\frac{q}{2}\varepsilon_{-}\tilde{\sigma}_2\,, \notag\\
\left\langle Q_{\alpha}^{+} R_{\beta}\right\rangle &=& 2\left(\varepsilon_{-}\tilde{\sigma}_2+\tilde{\sigma}_1\right)C_{\alpha\beta}\,, \notag\\
\left\langle Q_{\alpha}^{-} Q_{\beta}^{-}\right\rangle &=& 2\left(\varepsilon_{-}\tilde{\sigma}_2+\tilde{\sigma}_1\right)C_{\alpha\beta}\,,
\label{sMBIT2}
\end{eqnarray}
where the NR parameters are related to the relativistic $\mathcal{N}=2$ MB ones (see eqs.~\eqref{IT1} and \eqref{IT2}) as
\begin{equation}
    \tilde{\sigma}_1=\lambda_2\sigma_1\,, \qquad \tilde{\sigma}_2=\lambda_2\sigma_2\,.
\end{equation}
Then, the NR supergravity action based on the NR superalgebra given by eqs.~\eqref{sMBNR1}--\eqref{sMBNR3} is obtained by considering the components of the invariant tensor \eqref{sMBIT1}--\eqref{sMBIT2}, together with the gauge connection one-form \eqref{NRoneform}, in the CS action \eqref{CS}. In this way, after performing the redefinitions of the NR parameters,
\begin{align}
    \tilde\sigma_0&=p\tilde\sigma_1+q\tilde\sigma_3\,, & \tilde\sigma_3&=p\tilde\sigma_2+q\tilde\sigma_1\,, \label{redef}
\end{align}
we obtain
 \begin{eqnarray}
I_{\text{sNRMB}}=\int\tilde{\sigma}_{0}\tilde{L}_{0}+\tilde{\sigma}_{1}\tilde{L}_{1}+\tilde{\sigma}_{2} \tilde{L}_{2}+\tilde{\sigma}_3 \tilde{L}_{3}\,,\label{sNRMB}
\end{eqnarray}
with
\begin{align}
   \tilde{L}_0&= -\epsilon^{ac}\tau e_{a}e_{c}\,,\notag\\
     \tilde{L}_1&= e_{a}\hat{R}^{a}\left(\omega ^{b}\right)+\omega_{a}\hat{R}^{a}\left(e ^{b}\right)-2m \hat R(\omega )-2s \hat R(\tau)+2t_1 du_2+2t_2 du_1\notag \\
 &+2\left(\frac{q}{2}-\varepsilon_{-}\right)u_2du_1-2\bar{\psi}^{+}\nabla\rho-2\bar \rho\nabla\psi^{+}-2\bar{\psi}^{-}\nabla\psi^{-}\,, \notag \\
 \tilde{L}_2&=\omega _{a}\hat{R}^{a}\left(\omega
^{b}\right)-2sR\left( \omega \right) +2t_1dt_2-2q t_1du_2-2q t_2du_1+q\varepsilon_{-}u_2du_1 \notag \\
 &-2\varepsilon_{-}\bar{\psi}^{+}\nabla\rho-2\varepsilon_{-}\bar \rho\nabla\psi^{+}-2\varepsilon_{-}\bar{\psi}^{-}\nabla\psi^{-}\,, \notag \\
 \tilde{L}_3&=e_{a}\hat{R}^{a}\left(e ^{b}\right)-m\hat R(\tau )-\tau \hat{R}(m)\,.
\end{align}
Here we have defined the $(p,q)$-independent bosonic curvatures as
\begin{align}
 \hat{R}\left( \omega\right)  &=d\omega\,, \notag\\
 \hat{R}(\tau)&=d\tau\,, \notag\\
    \hat{R}^{a}\left( e ^{b}\right)  &= de ^{a}+\epsilon^{ac}\omega e_{c}+\epsilon^{ac}\tau \omega_{c}\,, \notag \\
     \hat{R}^{a}\left( \omega^{b}\right) &= d \omega^a + \epsilon ^{ac}\omega \omega_{c}\,,\notag\\
    \hat{R}\left( m\right) &= dm+ \epsilon ^{ac}e_{a} \omega_{c}\,. 
\end{align}
The action \eqref{sNRMB} defines the most general NR MB supergravity theory. As in the relativistic MB supergravity case, the fermionic contributions appear in the sectors proportional to $\tilde\sigma_1$ and $\tilde\sigma_2$. Interestingly, by performing the redefinitions of the NR parameters as in \eqref{redef}, the action can be written solely in terms of $\tilde\sigma_1$ and $\tilde\sigma_2$, which allows one to recover various known NR models as particular cases. In particular, for $p=0$ and $q=-2/\ell$, the action reduces to the NR analogue of the $\mathcal{N}=2$ teleparallel CS supergravity constructed in \cite{Concha:2021llq}. Conversely, setting $q=0$ and $p=1/\ell^2$, one recovers the NR counterpart of the $\mathcal{N}=2$ AdS supergravity, known as the extended Newton--Hooke CS supergravity theory \cite{Ozdemir:2019tby,Concha:2024dap}. Finally, in the limit $(p,q)\rightarrow0$, we recover the most general extended Bargmann
supergravity action \cite{Bergshoeff:2016lwr,deAzcarraga:2019mdn,Concha:2020eam}.

Let us conclude this section by noting that the consistency of the CS formulation requires the invariant tensor to be non-degenerate. 
This condition translates into constraints on the free parameters $\tilde{\sigma}_1$ and $\tilde{\sigma}_2$, ensuring that the pairing between the generators is invertible. In particular, non-degeneracy requires the following conditions:
\begin{align}
\tilde{\sigma}_1 &\neq \pm \varepsilon_{\pm}\tilde{\sigma}_2\,, &
\tilde{\sigma}_1 &\neq \frac{q}{2}\tilde{\sigma}_2\,, &
\tilde{\sigma}_1 &\neq \left(2q-\varepsilon_{-}\right)\tilde{\sigma}_2\,.
\end{align}
These conditions reproduce, as particular cases, the non-degeneracy requirements for the NR teleparallel, extended Newton--Hooke and extended Bargmann superalgebras, upon specifying the values of $p$ and $q$ as in Table~\ref{Table3}. 
Moreover, the non-degeneracy of the invariant tensor implies that the equations of motion are given by the vanishing of the full curvature two-form \eqref{supertorcuv},
\begin{equation}
    F = 0\,,
\end{equation}
which is equivalent to setting all bosonic and fermionic curvatures to zero. Of particular interest is the vanishing of $R(\omega)$, $R^{a}(\omega^{b})$ and $R^{a}(e^{b})$, which allows one to interpret the parameters $(p,q)$ as sources for the NR curvature $\hat{R}(\omega)$, the NR spatial curvature $\hat{R}^{a}(\omega^{b})$, and the spatial NR supertorsion $\hat{F}^{a}(e^{b})=\hat{R}^{a}(e^{b})+\bar{\psi}^+ \gamma^a \psi^-$:
\begin{align}
    \hat{R}(\omega)&=-\frac{\varepsilon_{-}}{2} \bar{\psi}^+ \gamma^0 \psi^+\,, \notag\\
    \hat{R}^{a}(\omega^{b})&=-p\,\epsilon^{ac}\tau e_{c}-\varepsilon_{-} \bar{\psi}^+ \gamma^a \psi^- \,,\notag\\
    \hat{F}^{a}(e^{b})&=-q\,\epsilon^{ac}\tau e_{c}\,.
\end{align}
Different choices of $(p,q)$ recover known NR supergravity models, extending the results obtained in NR MB gravity \cite{Concha:2023ejs} to the supersymmetric case.  For $p=0$ and $q=-2/\ell$, one obtains the NR teleparallel formulation \cite{Concha:2021llq}, where the cosmological constant acts as the source of spatial supertorsion. In the flat case $(p=q=0)$, both the NR curvatures and the spatial supertorsion vanish, reproducing the extended Bargmann supergravity dynamics \cite{Bergshoeff:2016lwr,deAzcarraga:2019mdn,Concha:2020eam}. For $p=1/\ell^{2}$ and $q=0$, one recovers the extended Newton--Hooke supergravity \cite{Ozdemir:2019tby,Concha:2024dap}, characterized by vanishing spatial supertorsion, with the cosmological constant acting as a source for the NR spatial supercurvature 
$\hat{F}^{a}(\omega^{b})=\hat{R}^{a}(\omega^{b})+\varepsilon_{-}\bar{\psi}^+ \gamma^a \psi^-$.

This structure highlights the role of the NR MB supergravity theory as a unifying framework that interpolates between different NR three-dimensional supergravity models.

\section{Conclusions}\label{sec5}
In this work, we have constructed a general class of three-dimensional NR CS supergravity theory that incorporates both curvature and torsion within the MB framework. The well-known difficulties in defining NR supergravity theories, namely the finiteness of the action, closure of the superalgebra, the existence of a non-degenerate invariant bilinear form, and the realization of the temporal translation generator as the anticommutator of two supercharges, are consistently resolved by starting from a $\mathfrak{so}(2)$ extension of the relativistic $\mathcal{N}=2$ MB superalgebra. We have shown that a proper NR formulation of the MB supergravity theory is naturally achieved through a NR expansion via the semigroup expansion method \cite{Concha:2023bly,Concha:2024dap}, rather than a naive contraction. The resulting superalgebra and supergravity action are characterized by the parameters $(p,q)$, which allow one to recover several known NR supergravity models, with or without torsion, for particular values. These results establish a consistent and systematic framework for NR supergravity with torsion in three dimensions.

Our construction opens several directions for future research. In particular, it would be interesting to study the asymptotic symmetries of the MB CS supergravity theory and to explore how the presence of non-vanishing torsion modifies the asymptotic symmetry algebra and conserved charges of standard AdS supergravity. One may ask whether the MB $(p,q)$ parameters can be fixed in such a way as to reproduce known asymptotic symmetry superalgebras, such as the conformal and $\mathfrak{bms}_3$ superalgebras \cite{Barnich:2014cwa,Barnich:2015sca,Caroca:2018obf}. It would also be natural to investigate their possible realization within the context of NR holography.

Another natural direction is the extension of our construction to the Carroll regime, where the inclusion of torsion in supersymmetric settings remains an open problem. It would be interesting to explore whether the methods developed here can be adapted to this limit, or alternatively to build upon recent Carrollian MB constructions \cite{Concha:2025vhd}. In this context, further analysis of the role of the $q$ parameter and the relativistic MB coupling constants may be required in order to avoid divergences in the ultra-relativistic limit.

Finally, the methodology employed here, together with its bosonic counterpart \cite{Concha:2023ejs}, paves the way for the construction of higher-spin extensions of MB CS gravity and, more generally, non-Lorentzian (super)gravity theories with extended symmetries. We expect that the framework developed in this work will provide new insights into the interplay between higher-spin symmetry, supersymmetry, torsion, and NR limits of gravity.

\section*{Acknowledgments}

P.C. and E.R. would like to thank the Instituto de Ciencias Exactas y Naturales (ICEN) of the Universidad Arturo Prat, Chile for hospitality. This work was funded by the National Agency for Research and Development ANID through FONDECYT grants No. 1231133, 1250642. This work was
supported by the Research project Code DIREG 04/2025 (P.C.) of the Universidad Católica de la Santísima Concepción (UCSC), Chile.  P.C. and E.R. would like to thank to the Dirección de Investigación and Vice-rectoría de Investigación of the Universidad Católica de la Santísima Concepción, Chile, for their constant support.

\appendix

\section{$\mathcal{N}$-extended Mielke-Baekler superalgebra}\label{appa}
An $\mathcal{N}$-extended Mielke--Baekler superalgebra with $\mathcal{N}=\mathfrak{p}+\mathfrak{q}$ is spanned by the set of generators
\begin{equation}
\{J_{A},P_{A},Z^{ij},Z^{IJ},Q^{i}_{\alpha},Q^{I}_{\alpha}\}\,,
\end{equation}
where $i=1,\dots,\mathfrak{p}$ and $I=1,\dots,\mathfrak{q}$. In addition to the bosonic MB generators, the algebra contains $\mathfrak{p}(\mathfrak{p}-1)/2+\mathfrak{q}(\mathfrak{q}-1)/2$ internal symmetry generators $Z^{ij}=-Z^{ji}$ and $Z^{IJ}=-Z^{JI}$. The non-vanishing (anti-)commutation relations are given by
\begin{eqnarray}
    \left[J_{A},J_{B}\right]&=&\epsilon_{ABC} J^{C}\,,\notag\\
    \left[J_{A},P_{B}\right]&=&\epsilon_{ABC} P^{C}\,,\notag\\
    \left[P_{A},P_{B}\right]&=&\epsilon_{ABC} \left(pJ^{C}+qP^{C}\right)\,,\notag\ \\
    \left[Z^{ij},Z^{kl}\right]&=&\delta^{jk}Z^{il}-\delta^{ik}Z^{jl}-\delta^{jl}Z^{ik}+\delta^{il}Z^{jk}\,,\notag\ \\
      \left[Z^{IJ},Z^{KL}\right]&=&\delta^{JK}Z^{IL}-\delta^{IK}Z^{JL}-\delta^{JL}Z^{IK}+\delta^{IL}Z^{JK}\,,\notag\ \\
\left[J_{A},Q_{\alpha}^{i}\right]&=&-\frac{1}{2}(\gamma_A)_{\alpha}^{\,\beta}Q_\beta^{i}\,,\notag\\
\left[J_{A},Q_{\alpha}^{I}\right]&=&-\frac{1}{2}(\gamma_A)_{\alpha}^{\,\beta}Q_\beta^{I}\,,\notag\\
\left[P_{A},Q_{\alpha}^{i}\right]&=&-\frac{\varepsilon_{+}}{2}(\gamma_A)_{\alpha}^{\,\beta}Q_\beta^{i}\,,\notag\\
\left[P_{A},Q_{\alpha}^{I}\right]&=&\frac{\varepsilon_{-}}{2}(\gamma_A)_{\alpha}^{\,\beta}Q_\beta^{I}\,,\notag\\
\left[Z^{ij},Q_{\alpha}^{k}\right]&=&  \delta^{jk}Q_{\alpha}^{i}-\delta^{ik}Q_{\alpha}^{j}\,,\notag \\
\left[Z^{IJ},Q_{\alpha}^{K}\right]&=&  \delta^{JK}Q_{\alpha}^{I}-\delta^{IK}Q_{\alpha}^{J}\,,\notag \\
\{Q_{\alpha}^{i},Q_{\beta}^{j}\}&=&-\delta^{ij
}\left(\gamma^{A}C\right)_{\alpha\beta}\left(\varepsilon_{-}J_A+P_A\right)-\varepsilon_{-}C_{\alpha\beta}Z^{ij}\,,\notag \\
\{Q_{\alpha}^{I},Q_{\beta}^{J}\}&=&-\delta^{IJ
}\left(\gamma^{A}C\right)_{\alpha\beta}\left(\varepsilon_{+}J_A+P_A\right)+\varepsilon_{-}C_{\alpha\beta}Z^{IJ}\,,\label{NMB}
\end{eqnarray}
Then, in order to recover the $(\mathfrak{p},\mathfrak{q})$ Poincaré superalgebra extended with the $\mathfrak{so}(\mathfrak{p})\oplus\mathfrak{so}(\mathfrak{q})$ automorphism algebra when the MB parameters $p,q$ are set to zero, we introduce additional bosonic generators $U^{ij}=-U^{ji}$ and $U^{IJ}=-U^{JI}$ satisfying
\begin{eqnarray}
\left[U^{ij},U^{kl}\right]&=&-\varepsilon_{-}\left(\delta^{jk}U^{il}-\delta^{ik}U^{jl}-\delta^{jl}U^{ik}+\delta^{il}U^{jk}\right)\,,\notag\ \\
\left[U^{IJ},U^{KL}\right]&=&-\varepsilon_{-}\left(\delta^{JK}U^{IL}-\delta^{IK}U^{JL}-\delta^{JL}U^{IK}+\delta^{IL}U^{JK}\right)\,.\label{UU}
\end{eqnarray}
Introducing the redefinition,
\begin{equation}
    T^{ij}=Z^{ij}-\frac{1}{\varepsilon_{-}}U^{ij}\,, \qquad T^{IJ}=Z^{IJ}-\frac{1}{\varepsilon_{-}}U^{IJ}\,,
\end{equation}
one can eliminate the generators $Z^{ij}$ and $Z^{IJ}$ and obtain a well-defined flat limit $(p,q)\to 0$. 
With this redefinition, the resulting superalgebra, defined as the direct sum of the $(\mathfrak{p},\mathfrak{q})$ MB superalgebra and the $\mathfrak{so}(\mathfrak{p})\oplus\mathfrak{so}(\mathfrak{q})$ automorphism algebra, satisfies the MB algebra \eqref{MB}, together with \eqref{UU} and
\begin{eqnarray}
\left[T^{ij},T^{kl}\right]&=&\delta^{jk}T^{il}-\delta^{ik}T^{jl}-\delta^{jl}T^{ik}+\delta^{il}T^{jk}\,,\notag \\
\left[T^{IJ},T^{KL}\right]&=&\delta^{JK}T^{IL}-\delta^{IK}T^{JL}-\delta^{JL}T^{IK}+\delta^{IL}T^{JK}\,,\notag\\
\left[T^{ij},U^{kl}\right]&=&\delta^{jk}U^{il}-\delta^{ik}U^{jl}-\delta^{jl}U^{ik}+\delta^{il}U^{jk}\,,\notag\ \\
\left[T^{IJ},U^{KL}\right]&=&\delta^{JK}U^{IL}-\delta^{IK}U^{JL}-\delta^{JL}U^{IK}+\delta^{IL}U^{JK}\,,\notag\\
\left[J_{A},Q_{\alpha}^{i}\right]&=&-\frac{1}{2}(\gamma_A)_{\alpha}^{\,\beta}Q_\beta^{i}\,,\notag\\
\left[J_{A},Q_{\alpha}^{I}\right]&=&-\frac{1}{2}(\gamma_A)_{\alpha}^{\,\beta}Q_\beta^{I}\,,\notag\\
\left[P_{A},Q_{\alpha}^{i}\right]&=&-\frac{\varepsilon_{+}}{2}(\gamma_A)_{\alpha}^{\,\beta}Q_\beta^{i}\,,\notag\\
\left[P_{A},Q_{\alpha}^{I}\right]&=&\frac{\varepsilon_{-}}{2}(\gamma_A)_{\alpha}^{\,\beta}Q_\beta^{I}\,,\notag\\
\left[T^{ij},Q_{\alpha}^{k}\right]&=&  \delta^{jk}Q_{\alpha}^{i}-\delta^{ik}Q_{\alpha}^{j}\,,\notag \\
\left[T^{IJ},Q_{\alpha}^{K}\right]&=&  \delta^{JK}Q_{\alpha}^{I}-\delta^{IK}Q_{\alpha}^{J}\,,\notag \\
\{Q_{\alpha}^{i},Q_{\beta}^{j}\}&=&-\delta^{ij
}\left(\gamma^{A}C\right)_{\alpha\beta}\left(\varepsilon_{-}J_A+P_A\right)-C_{\alpha\beta}\left(\varepsilon_{-}T^{ij}+U^{ij}\right)\,,\notag \\
\{Q_{\alpha}^{I},Q_{\beta}^{J}\}&=&-\delta^{IJ
}\left(\gamma^{A}C\right)_{\alpha\beta}\left(\varepsilon_{+}J_A+P_A\right)+C_{\alpha\beta}\left(\varepsilon_{-}T^{IJ}+U^{IJ}\right)\,.\label{pqauto}
\end{eqnarray}
In analogy with the $(\mathfrak{p},\mathfrak{q})$ AdS supergravity case \cite{Howe:1995zm}, the direct sum of the $\mathfrak{so}(\mathfrak{p})\oplus\mathfrak{so}(\mathfrak{q})$ algebra and the $\mathcal{N}$-extended MB superalgebra leads to a non-degenerate extended Poincaré superalgebra in the flat limit $(p,q)\to 0$. The $\mathfrak{so}(\mathfrak{p})\oplus\mathfrak{so}(\mathfrak{q})$ extension of the $\mathcal{N}$-extended MB superalgebra admits the invariant tensor of the bosonic MB algebra \eqref{IT1}, together with the following non-vanishing components:
\begin{eqnarray}
    \langle {T}^{ij} {T}^{kl} \rangle &=& {\sigma}_2 \left(\delta^{il}\delta^{kj}-\delta^{ik}\delta^{lj}\right)\,, \notag \\
    \langle {T}^{IJ} {T}^{KL} \rangle &=& {\sigma}_2 \left(\delta^{IL}\delta^{KJ}-\delta^{IK}\delta^{LJ}\right)\,, \notag \\
    \langle {T}^{ij} {U}^{kl} \rangle &=& -\left(\sigma_1-q{\sigma}_2 \right)\left(\delta^{il}\delta^{kj}-\delta^{ik}\delta^{lj}\right)\,, \notag \\
    \langle {T}^{IJ} {U}^{KL} \rangle &=& \left(\sigma_1-q{\sigma}_2 \right) \left(\delta^{IL}\delta^{KJ}-\delta^{IK}\delta^{LJ}\right)\,, \notag \\
    \langle {U}^{ij} {U}^{kl} \rangle &=& -\left[\left(\frac{q}{2}-\varepsilon_{-}\right)\sigma_1+\frac{q}{2}\varepsilon_{-}\sigma_2 \right]\left(\delta^{il}\delta^{kj}-\delta^{ik}\delta^{lj}\right)\,, \notag \\
    \langle {U}^{IJ} {U}^{KL} \rangle &=& \left[\left(\frac{q}{2}-\varepsilon_{-}\right)\sigma_1+\frac{q}{2}\varepsilon_{-}\sigma_2 \right] \left(\delta^{IL}\delta^{KJ}-\delta^{IK}\delta^{LJ}\right)\,, \notag \\
     \langle Q_{\alpha}^{i},Q_{\beta}^{j}\rangle &=& 2\left(\varepsilon_{-}\sigma_2+\sigma_1\right)C_{\alpha\beta}\delta^{ij}\,,\notag\\
      \langle Q_{\alpha},Q_{\beta}\rangle &=& 2\left(-\varepsilon_{-}\sigma_2+\sigma_1\right)C_{\alpha\beta}\delta^{IJ}\,. \label{IT2}
\end{eqnarray}

\bibliographystyle{fullsort.bst}
 
\bibliography{Non_relativistic_MB_Supergravity}

\providecommand{\href}[2]{#2}\begingroup\raggedright\begin{thebibliography}{10}

\bibitem{Banados:1992wn}
M.~Banados, C.~Teitelboim, and J.~Zanelli, ``{The Black hole in
  three-dimensional space-time},'' {\em Phys. Rev. Lett.} {\bf 69} (1992)
  1849--1851, \href{http://www.arXiv.org/abs/hep-th/9204099}{{\tt
  hep-th/9204099}}.

\bibitem{Mielke:1991nn}
E.~W. Mielke and P.~Baekler, ``{Topological gauge model of gravity with
  torsion},'' {\em Phys. Lett. A} {\bf 156} (1991) 399--403.

\bibitem{Geiller:2020edh}
M.~Geiller, C.~Goeller, and N.~Merino, ``{Most general theory of 3d gravity:
  Covariant phase space, dual diffeomorphisms, and more},'' {\em JHEP} {\bf 02}
  (2021) 120, \href{http://www.arXiv.org/abs/2011.09873}{{\tt 2011.09873}}.

\bibitem{Blagojevic:2003uc}
M.~Blagojevic and M.~Vasilic, ``{Asymptotic symmetries in 3-d gravity with
  torsion},'' {\em Phys. Rev. D} {\bf 67} (2003) 084032,
  \href{http://www.arXiv.org/abs/gr-qc/0301051}{{\tt gr-qc/0301051}}.

\bibitem{Blagojevic:2003vn}
M.~Blagojevic and M.~Vasilic, ``{3-D gravity with torsion as a Chern-Simons
  gauge theory},'' {\em Phys. Rev. D} {\bf 68} (2003) 104023,
  \href{http://www.arXiv.org/abs/gr-qc/0307078}{{\tt gr-qc/0307078}}.

\bibitem{Blagojevic:2006hh}
M.~Blagojevic and B.~Cvetkovic, ``{Black hole entropy from the boundary
  conformal structure in 3D gravity with torsion},'' {\em JHEP} {\bf 10} (2006)
  005, \href{http://www.arXiv.org/abs/gr-qc/0606086}{{\tt gr-qc/0606086}}.

\bibitem{Giacomini:2006dr}
A.~Giacomini, R.~Troncoso, and S.~Willison, ``{Three-dimensional supergravity
  reloaded},'' {\em Class. Quant. Grav.} {\bf 24} (2007) 2845--2860,
  \href{http://www.arXiv.org/abs/hep-th/0610077}{{\tt hep-th/0610077}}.

\bibitem{Cvetkovic:2007sr}
B.~Cvetkovic and M.~Blagojevic, ``{Supersymmetric 3D gravity with torsion:
  Asymptotic symmetries},'' {\em Class. Quant. Grav.} {\bf 24} (2007)
  3933--3950, \href{http://www.arXiv.org/abs/gr-qc/0702121}{{\tt
  gr-qc/0702121}}.

\bibitem{Klemm:2007yu}
D.~Klemm and G.~Tagliabue, ``{The CFT dual of AdS gravity with torsion},'' {\em
  Class. Quant. Grav.} {\bf 25} (2008) 035011,
  \href{http://www.arXiv.org/abs/0705.3320}{{\tt 0705.3320}}.

\bibitem{Blagojevic:2013bu}
M.~Blagojevic, B.~Cvetkovic, O.~Miskovic, and R.~Olea, ``{Holography in 3D AdS
  gravity with torsion},'' {\em JHEP} {\bf 05} (2013) 103,
  \href{http://www.arXiv.org/abs/1301.1237}{{\tt 1301.1237}}.

\bibitem{Peleteiro:2020ubv}
J.~Peleteiro and C.~Valc\'arcel, ``{Spin-3 fields in Mielke-Baekler gravity},''
  {\em Class. Quant. Grav.} {\bf 37} (2020), no.~18, 185010,
  \href{http://www.arXiv.org/abs/2003.02627}{{\tt 2003.02627}}.

\bibitem{Son:2008ye}
D.~Son, ``{Toward an AdS/cold atoms correspondence: A Geometric realization of
  the Schrodinger symmetry},'' {\em Phys. Rev. D} {\bf 78} (2008) 046003,
  \href{http://www.arXiv.org/abs/0804.3972}{{\tt 0804.3972}}.

\bibitem{Balasubramanian:2008dm}
K.~Balasubramanian and J.~McGreevy, ``{Gravity duals for non-relativistic
  CFTs},'' {\em Phys. Rev. Lett.} {\bf 101} (2008) 061601,
  \href{http://www.arXiv.org/abs/0804.4053}{{\tt 0804.4053}}.

\bibitem{Kachru:2008yh}
S.~Kachru, X.~Liu, and M.~Mulligan, ``{Gravity duals of Lifshitz-like fixed
  points},'' {\em Phys. Rev. D} {\bf 78} (2008) 106005,
  \href{http://www.arXiv.org/abs/0808.1725}{{\tt 0808.1725}}.

\bibitem{Taylor:2008tg}
M.~Taylor, ``{Non-relativistic holography},''
  \href{http://www.arXiv.org/abs/0812.0530}{{\tt 0812.0530}}.

\bibitem{Duval:2008jg}
C.~Duval, M.~Hassaine, and P.~A. Horvathy, ``{The Geometry of Schrodinger
  symmetry in gravity background/non-relativistic CFT},'' {\em Annals Phys.}
  {\bf 324} (2009) 1158--1167, \href{http://www.arXiv.org/abs/0809.3128}{{\tt
  0809.3128}}.

\bibitem{Bagchi:2009my}
A.~Bagchi and R.~Gopakumar, ``{Galilean Conformal Algebras and AdS/CFT},'' {\em
  JHEP} {\bf 07} (2009) 037, \href{http://www.arXiv.org/abs/0902.1385}{{\tt
  0902.1385}}.

\bibitem{Hartnoll:2009sz}
S.~A. Hartnoll, ``{Lectures on holographic methods for condensed matter
  physics},'' {\em Class. Quant. Grav.} {\bf 26} (2009) 224002,
  \href{http://www.arXiv.org/abs/0903.3246}{{\tt 0903.3246}}.

\bibitem{Bagchi:2009pe}
A.~Bagchi, R.~Gopakumar, I.~Mandal, and A.~Miwa, ``{GCA in 2d},'' {\em JHEP}
  {\bf 08} (2010) 004, \href{http://www.arXiv.org/abs/0912.1090}{{\tt
  0912.1090}}.

\bibitem{Hoyos:2011ez}
C.~Hoyos and D.~T. Son, ``{Hall Viscosity and Electromagnetic Response},'' {\em
  Phys. Rev. Lett.} {\bf 108} (2012) 066805,
  \href{http://www.arXiv.org/abs/1109.2651}{{\tt 1109.2651}}.

\bibitem{Son:2013rqa}
D.~T. Son, ``{Newton-Cartan Geometry and the Quantum Hall Effect},''
  \href{http://www.arXiv.org/abs/1306.0638}{{\tt 1306.0638}}.

\bibitem{Christensen:2013lma}
M.~H. Christensen, J.~Hartong, N.~A. Obers, and B.~Rollier, ``{Torsional
  Newton-Cartan Geometry and Lifshitz Holography},'' {\em Phys. Rev. D} {\bf
  89} (2014) 061901, \href{http://www.arXiv.org/abs/1311.4794}{{\tt
  1311.4794}}.

\bibitem{Christensen:2013rfa}
M.~H. Christensen, J.~Hartong, N.~A. Obers, and B.~Rollier, ``{Boundary
  Stress-Energy Tensor and Newton-Cartan Geometry in Lifshitz Holography},''
  {\em JHEP} {\bf 01} (2014) 057,
  \href{http://www.arXiv.org/abs/1311.6471}{{\tt 1311.6471}}.

\bibitem{Abanov:2014ula}
A.~G. Abanov and A.~Gromov, ``{Electromagnetic and gravitational responses of
  two-dimensional noninteracting electrons in a background magnetic field},''
  {\em Phys. Rev. B} {\bf 90} (2014), no.~1, 014435,
  \href{http://www.arXiv.org/abs/1401.3703}{{\tt 1401.3703}}.

\bibitem{Hartong:2014oma}
J.~Hartong, E.~Kiritsis, and N.~A. Obers, ``{Lifshitz space--times for
  Schrödinger holography},'' {\em Phys. Lett. B} {\bf 746} (2015) 318--324,
  \href{http://www.arXiv.org/abs/1409.1519}{{\tt 1409.1519}}.

\bibitem{Hartong:2014pma}
J.~Hartong, E.~Kiritsis, and N.~A. Obers, ``{Schrödinger Invariance from
  Lifshitz Isometries in Holography and Field Theory},'' {\em Phys. Rev. D}
  {\bf 92} (2015) 066003, \href{http://www.arXiv.org/abs/1409.1522}{{\tt
  1409.1522}}.

\bibitem{Hartong:2015wxa}
J.~Hartong, E.~Kiritsis, and N.~A. Obers, ``{Field Theory on Newton-Cartan
  Backgrounds and Symmetries of the Lifshitz Vacuum},'' {\em JHEP} {\bf 08}
  (2015) 006, \href{http://www.arXiv.org/abs/1502.00228}{{\tt 1502.00228}}.

\bibitem{Geracie:2015dea}
M.~Geracie, K.~Prabhu, and M.~M. Roberts, ``{Curved non-relativistic
  spacetimes, Newtonian gravitation and massive matter},'' {\em J. Math. Phys.}
  {\bf 56} (2015), no.~10, 103505,
  \href{http://www.arXiv.org/abs/1503.02682}{{\tt 1503.02682}}.

\bibitem{Gromov:2015fda}
A.~Gromov, K.~Jensen, and A.~G. Abanov, ``{Boundary effective action for
  quantum Hall states},'' {\em Phys. Rev. Lett.} {\bf 116} (2016), no.~12,
  126802, \href{http://www.arXiv.org/abs/1506.07171}{{\tt 1506.07171}}.

\bibitem{Hartong:2015zia}
J.~Hartong and N.~A. Obers, ``{Ho\v{r}ava-Lifshitz gravity from dynamical
  Newton-Cartan geometry},'' {\em JHEP} {\bf 07} (2015) 155,
  \href{http://www.arXiv.org/abs/1504.07461}{{\tt 1504.07461}}.

\bibitem{Taylor:2015glc}
M.~Taylor, ``{Lifshitz holography},'' {\em Class. Quant. Grav.} {\bf 33}
  (2016), no.~3, 033001, \href{http://www.arXiv.org/abs/1512.03554}{{\tt
  1512.03554}}.

\bibitem{Zaanen:2015oix}
J.~Zaanen, Y.-W. Sun, Y.~Liu, and K.~Schalm, {\em {Holographic Duality in
  Condensed Matter Physics}}.
\newblock Cambridge Univ. Press, 2015.

\bibitem{Devecioglu:2018apj}
D.~O. Devecioglu, N.~Ozdemir, M.~Ozkan, and U.~Zorba, ``{Scale invariance in
  Newton\textendash{}Cartan and Ho\v{r}ava\textendash{}Lifshitz gravity},''
  {\em Class. Quant. Grav.} {\bf 35} (2018), no.~11, 115016,
  \href{http://www.arXiv.org/abs/1801.08726}{{\tt 1801.08726}}.

\bibitem{Concha:2023ejs}
P.~Concha, N.~Merino, and E.~Rodr{\'\i}guez, ``{Non-relativistic limit of the
  Mielke{\textendash}Baekler gravity theory},'' {\em Eur. Phys. J. C} {\bf 84}
  (2024), no.~4, 407, \href{http://www.arXiv.org/abs/2309.00500}{{\tt
  2309.00500}}.

\bibitem{Bergshoeff:2014uea}
E.~A. Bergshoeff, J.~Hartong, and J.~Rosseel, ``{Torsional
  Newton\textendash{}Cartan geometry and the Schr\"odinger algebra},'' {\em
  Class. Quant. Grav.} {\bf 32} (2015), no.~13, 135017,
  \href{http://www.arXiv.org/abs/1409.5555}{{\tt 1409.5555}}.

\bibitem{Papageorgiou:2010ud}
G.~Papageorgiou and B.~J. Schroers, ``{Galilean quantum gravity with
  cosmological constant and the extended $q$-Heisenberg algebra},'' {\em JHEP}
  {\bf 11} (2010) 020, \href{http://www.arXiv.org/abs/1008.0279}{{\tt
  1008.0279}}.

\bibitem{Duval:2011mi}
C.~Duval and P.~Horvathy, ``{Conformal Galilei groups, Veronese curves, and
  Newton-Hooke spacetimes},'' {\em J. Phys. A} {\bf 44} (2011) 335203,
  \href{http://www.arXiv.org/abs/1104.1502}{{\tt 1104.1502}}.

\bibitem{Hartong:2016yrf}
J.~Hartong, Y.~Lei, and N.~A. Obers, ``{Nonrelativistic Chern-Simons theories
  and three-dimensional Ho\v rava-Lifshitz gravity},'' {\em Phys. Rev. D} {\bf
  94} (2016), no.~6, 065027, \href{http://www.arXiv.org/abs/1604.08054}{{\tt
  1604.08054}}.

\bibitem{Duval:2016tzi}
C.~Duval, G.~Gibbons, and P.~Horvathy, ``{Conformal and projective symmetries
  in Newtonian cosmology},'' {\em J. Geom. Phys.} {\bf 112} (2017) 197--209,
  \href{http://www.arXiv.org/abs/1605.00231}{{\tt 1605.00231}}.

\bibitem{Papageorgiou:2009zc}
G.~Papageorgiou and B.~J. Schroers, ``{A Chern-Simons approach to Galilean
  quantum gravity in 2+1 dimensions},'' {\em JHEP} {\bf 11} (2009) 009,
  \href{http://www.arXiv.org/abs/0907.2880}{{\tt 0907.2880}}.

\bibitem{Bergshoeff:2016lwr}
E.~A. Bergshoeff and J.~Rosseel, ``{Three-Dimensional Extended Bargmann
  Supergravity},'' {\em Phys. Rev. Lett.} {\bf 116} (2016), no.~25, 251601,
  \href{http://www.arXiv.org/abs/1604.08042}{{\tt 1604.08042}}.

\bibitem{Concha:2021llq}
P.~Concha, L.~Ravera, and E.~Rodr\'\i{}guez, ``{Three-dimensional
  non-relativistic supergravity and torsion},'' {\em Eur. Phys. J. C} {\bf 82}
  (2022), no.~3, 220, \href{http://www.arXiv.org/abs/2112.05902}{{\tt
  2112.05902}}.

\bibitem{Bergshoeff:2022iyb}
E.~A. Bergshoeff and J.~Rosseel, {\em {Non-Lorentzian Supergravity}}.
\newblock 2023.
\newblock \href{http://www.arXiv.org/abs/2211.02604}{{\tt 2211.02604}}.

\bibitem{Hatsuda:2001pp}
M.~Hatsuda and M.~Sakaguchi, ``{Wess-Zumino term for the AdS superstring and
  generalized Inonu-Wigner contraction},'' {\em Prog. Theor. Phys.} {\bf 109}
  (2003) 853--867, \href{http://www.arXiv.org/abs/hep-th/0106114}{{\tt
  hep-th/0106114}}.

\bibitem{deAzcarraga:2002xi}
J.~A. de~Azcarraga, J.~M. Izquierdo, M.~Picon, and O.~Varela, ``{Generating Lie
  and gauge free differential (super)algebras by expanding Maurer-Cartan forms
  and Chern-Simons supergravity},'' {\em Nucl. Phys. B} {\bf 662} (2003)
  185--219, \href{http://www.arXiv.org/abs/hep-th/0212347}{{\tt
  hep-th/0212347}}.

\bibitem{Izaurieta:2006zz}
F.~Izaurieta, E.~Rodriguez, and P.~Salgado, ``{Expanding Lie (super)algebras
  through Abelian semigroups},'' {\em J. Math. Phys.} {\bf 47} (2006) 123512,
  \href{http://www.arXiv.org/abs/hep-th/0606215}{{\tt hep-th/0606215}}.

\bibitem{deAzcarraga:2007et}
J.~de~Azcarraga, J.~Izquierdo, M.~Picon, and O.~Varela, ``{Expansions of
  algebras and superalgebras and some applications},'' {\em Int. J. Theor.
  Phys.} {\bf 46} (2007) 2738--2752,
  \href{http://www.arXiv.org/abs/hep-th/0703017}{{\tt hep-th/0703017}}.

\bibitem{Concha:2024dap}
P.~Concha and L.~Ravera, ``{Non-Lorentzian supergravity and kinematical
  superalgebras},'' {\em JHEP} {\bf 03} (2025) 032,
  \href{http://www.arXiv.org/abs/2412.07665}{{\tt 2412.07665}}.

\bibitem{deAzcarraga:2019mdn}
J.~A. de~Azc\'arraga, D.~G\'utiez, and J.~M. Izquierdo, ``{Extended $D = 3$
  Bargmann supergravity from a Lie algebra expansion},'' {\em Nucl. Phys. B}
  {\bf 946} (2019) 114706, \href{http://www.arXiv.org/abs/1904.12786}{{\tt
  1904.12786}}.

\bibitem{Ozdemir:2019tby}
N.~Ozdemir, M.~Ozkan, and U.~Zorba, ``{Three-dimensional extended Lifshitz,
  Schrödinger and Newton-Hooke supergravity},'' {\em JHEP} {\bf 11} (2019)
  052, \href{http://www.arXiv.org/abs/1909.10745}{{\tt 1909.10745}}.

\bibitem{Concha:2019mxx}
P.~Concha, L.~Ravera, and E.~Rodríguez, ``{Three-dimensional Maxwellian
  extended Bargmann supergravity},'' {\em JHEP} {\bf 04} (2020) 051,
  \href{http://www.arXiv.org/abs/1912.09477}{{\tt 1912.09477}}.

\bibitem{Concha:2020tqx}
P.~Concha, L.~Ravera, and E.~Rodr\'\i{}guez, ``{Three-dimensional
  non-relativistic extended supergravity with cosmological constant},'' {\em
  Eur. Phys. J. C} {\bf 80} (2020), no.~12, 1105,
  \href{http://www.arXiv.org/abs/2008.08655}{{\tt 2008.08655}}.

\bibitem{Concha:2021jos}
P.~Concha, L.~Ravera, and E.~Rodr\'\i{}guez, ``{Three-dimensional exotic
  Newtonian supergravity theory with cosmological constant},'' {\em Eur. Phys.
  J. C} {\bf 81} (2021), no.~7, 646,
  \href{http://www.arXiv.org/abs/2104.12908}{{\tt 2104.12908}}.

\bibitem{Concha:2024vql}
P.~Concha, E.~Rodr{\'\i}guez, and S.~Salgado, ``{Three-dimensional
  non-relativistic Hietarinta supergravity},'' {\em Eur. Phys. J. C} {\bf 85}
  (2025), no.~1, 47, \href{http://www.arXiv.org/abs/2409.01298}{{\tt
  2409.01298}}.

\bibitem{Concha:2020eam}
P.~Concha, M.~Ipinza, L.~Ravera, and E.~Rodr\'\i{}guez, ``{Non-relativistic
  three-dimensional supergravity theories and semigroup expansion method},''
  {\em JHEP} {\bf 02} (2021) 094,
  \href{http://www.arXiv.org/abs/2010.01216}{{\tt 2010.01216}}.

\bibitem{Caroca:2021njq}
R.~Caroca, P.~Concha, D.~Pe\~nafiel, and E.~Rodr\'\i{}guez,
  ``{Three-dimensional teleparallel Chern-Simons supergravity theory},'' {\em
  Eur. Phys. J. C} {\bf 81} (2021), no.~8, 762,
  \href{http://www.arXiv.org/abs/2103.06717}{{\tt 2103.06717}}.

\bibitem{Witten:1988hc}
E.~Witten, ``{(2+1)-Dimensional Gravity as an Exactly Soluble System},'' {\em
  Nucl. Phys. B} {\bf 311} (1988) 46.

\bibitem{Zanelli:2005sa}
J.~Zanelli, ``{Lecture notes on Chern-Simons (super-)gravities. Second edition
  (February 2008)},'' in {\em {7th Mexican Workshop on Particles and Fields}}.
\newblock 2, 2005.
\newblock \href{http://www.arXiv.org/abs/hep-th/0502193}{{\tt hep-th/0502193}}.

\bibitem{Concha:2025vhd}
P.~Concha, N.~Merino, L.~Ravera, and E.~Rodr{\'\i}guez, ``{Torsional Carroll
  Gravity},'' {\em Phys. Rev. Lett.} {\bf 136} (2026), no.~10, 101402,
  \href{http://www.arXiv.org/abs/2512.14688}{{\tt 2512.14688}}.

\bibitem{Matulich:2019cdo}
J.~Matulich, S.~Prohazka, and J.~Salzer, ``{Limits of three-dimensional gravity
  and metric kinematical Lie algebras in any dimension},'' {\em JHEP} {\bf 07}
  (2019) 118, \href{http://www.arXiv.org/abs/1903.09165}{{\tt 1903.09165}}.

\bibitem{Concha:2023bly}
P.~Concha, D.~Pino, L.~Ravera, and E.~Rodr{\'\i}guez, ``{Extended kinematical
  3D gravity theories},'' {\em JHEP} {\bf 01} (2024) 040,
  \href{http://www.arXiv.org/abs/2310.01335}{{\tt 2310.01335}}.

\bibitem{Barnich:2014cwa}
G.~Barnich, L.~Donnay, J.~Matulich, and R.~Troncoso, ``{Asymptotic symmetries
  and dynamics of three-dimensional flat supergravity},'' {\em JHEP} {\bf 08}
  (2014) 071, \href{http://www.arXiv.org/abs/1407.4275}{{\tt 1407.4275}}.

\bibitem{Barnich:2015sca}
G.~Barnich, L.~Donnay, J.~Matulich, and R.~Troncoso, ``{Super-BMS$_{3}$
  invariant boundary theory from three-dimensional flat supergravity},'' {\em
  JHEP} {\bf 01} (2017) 029, \href{http://www.arXiv.org/abs/1510.08824}{{\tt
  1510.08824}}.

\bibitem{Caroca:2018obf}
R.~Caroca, P.~Concha, O.~Fierro, and E.~Rodr\'\i{}guez, ``{Three-dimensional
  Poincar\'e supergravity and $N$-extended supersymmetric $BMS_3$ algebra},''
  {\em Phys. Lett. B} {\bf 792} (2019) 93--100,
  \href{http://www.arXiv.org/abs/1812.05065}{{\tt 1812.05065}}.

\bibitem{Howe:1995zm}
P.~S. Howe, J.~Izquierdo, G.~Papadopoulos, and P.~Townsend, ``{New
  supergravities with central charges and Killing spinors in
  (2+1)-dimensions},'' {\em Nucl. Phys. B} {\bf 467} (1996) 183--214,
  \href{http://www.arXiv.org/abs/hep-th/9505032}{{\tt hep-th/9505032}}.

\end{thebibliography}\endgroup

\end{document}